\documentstyle[11pt,aaspp4,tighten]{article}
\def\rosat{{\sl ROSAT~}}
\def\asca{{\sl ASCA~}}

\slugcomment{}
\begin{document}

\title{\bf \rosat and \asca Observations of the Crab-Like Supernova Remnant
N157B in the Large Magellanic Cloud}

\author{Q. Daniel Wang}
\affil{Dearborn Observatory, Northwestern University}
\affil{ 2131 Sheridan Road, Evanston,~IL 60208-2900}
\affil{Electronic mail: wqd@nwu.edu}
\affil{and}
\author{Eric V. Gotthelf}
\affil{NASA/Goddard Space Flight Center}
\affil{Greenbelt, MD 20771}
\affil{Electronic mail: gotthelf@gsfc.nasa.gov}

\begin{abstract}

	We report the results of \rosat and \asca X-ray observations of 
the supernova remnant N157B (or 30 Dor B, SNR 0539-69.1) in the Large 
Magellanic Cloud. For comparison, we also briefly describe the results
on SNR 0540-69.3, the only confirmed Crab-like remnant in the Cloud. 
The X-ray emission from N157B can be decomposed into a bright 
comet-shaped feature, superimposed on a diffuse emission region of a dimension 
$\sim 20$~pc. The flat and {\sl nearly} featureless spectrum of the remnant
is distinctly different from those of young 
shell-like remnants, suggesting a predominantly Crab-like nature of N157B.
Characterized by a power law with an energy slope $\sim 1.5$, the spectrum of 
N157B above $\sim 2$~keV is, however, considerably steeper than that of 
SNR 0540-69.3, which has a slope of $\sim 1.0$. At lower energies, the 
spectrum of N157B presents marginal evidence for emission lines, which if real 
most likely arise in hot gas of the diffuse emission region. 
The hot gas has a characteristic thermal temperature of 0.4-0.7~keV. 
No significant periodic signal 
is detected from N157B in the period range of $3 \times 10^{-3}-2000$~s.
The pulsed fraction is $\lesssim 9\%$ (99\% confidence) in the $2-7$~keV 
range. 

	We discuss the nature of the individual X-ray components. 
In particular, we suggest that 
the synchrotron radiation of relativistic particles from a 
fast-moving ($\sim 10^3 {\rm~km~s^{-1}}$) pulsar 
explains the size, morphology, spectrum, and energetics of the 
comet-shaped X-ray feature. We infer the age of the remnant as 
$\sim 5 \times 10^3$~yrs. The lack of radio polarization 
of the remnant may be due to Faraday dispersion by foreground \ion{H}{2} gas. 

\end{abstract}

\section {Introduction}

	Only three supernova remnants (SNRs) have been unambiguously confirmed 
to be Crab-like: SNR 0540-69.3, MSH 15-52, and the
Crab nebula itself. These remnants show centrally peaked morphologies and flat 
emission spectra; the emission comes predominantly from synchrotron 
nebulae powered by embedded young pulsars. There are a number of SNRs
that have similar morphologies and spectral characteristics, but contain 
no detected pulsars. N157B is such a remnant (Mills, Turtle, \& 
Watkinson 1978; Clark et al. 1982; 
Mathewson et al. 1983; Wang \& Helfand 1991; Chu et al. 1992 ---
CKSL92 hereafter; Dickel et al. 1994 --- D94 hereafter). 
Like SNR 0540-69.3, N157B is in the Large Magellanic Cloud (LMC) 
and thus has a good distance determination. But,
previous observations have shown no clear correspondence between the
morphologies of N157B in radio, optical, and X-ray. 
The remnant's age of $\sim 2.4 \times 10^4$~yrs, inferred from kinematics 
of optical features (CKSL92) is also much greater
(by a factor of about 10) than the confirmed Crab-like SNRs. 
No radio polarized emission has been detected from N157B,
as might be expected for a synchrotron nebula (D94). 
In short, the exact nature of the remnant has been uncertain.

	Here we report new results on N157B, based primarily on observations 
from the \rosat and \asca X-ray Observatories. 
These observations provide much improved timing, imaging, and spectral 
resolution capabilities. For comparison, we also present the results of 
analyzing a 40~ks \asca observation on SNR 0540-69.3. 
We describe the reduction of the X-ray observations in \S2, 
and present our analysis and results in \S3. Then in \S4, we discuss 
the implications of our results, together with
previous observations, and postulate 
scenarios to explain various components of N157B.  
We summarize our results and conclusions in \S5. Appendix A 
describes an algorithm we used for extracting \asca spectra with correction
for the instrument point spread function (PSF). 
Throughout the paper, we adopt
an LMC distance $D \approx 47$~kpc (Gould 1995), at which $1^\prime$ 
corresponds to $14$~pc.

\section {Observations and Data Reduction}

\subsection {\asca Observations}

	Table 1 summarizes parameters of the two \asca observations used
in this study. Both observations were acquired during {\sl ASCA}'s 
Performance Verification phase, and the data were made available through 
the public archive. Since the \asca has two types of
instruments, which observe simultaneously, a pair of Solid-state Imaging
Spectrometers (SIS) and a pair of Gas Imaging Spectrometers (GIS), the two
\asca observations yielded eight sets of data: four from the SISs and
four from the GISs.  The SIS has an energy coverage approximately
in the 0.5-8~keV range, and the GIS in the 0.8-10~keV range
(Tanaka, Inoue, \& Holt 1994). 

Data from the SIS detectors were collected in 4-CCD FAINT and BRIGHT modes.
The data were processed by converting the FAINT mode data 
to BRIGHT mode equivalent data and by filtering the combined set through the 
standard cleaning criteria. CCD pixels with anomalous counts were identified 
and excluded using the CLEANSIS technique (Gotthelf 1993). 
Detailed procedures of these methods
can be found in Day et al. (1995). Fig.~1 presents an exposure corrected, 
smoothed SIS image of the 30 Dor region created by combining screened data 
from both observations. 

	The SIS data are the most useful in our spectral study of N157B, 
because of both the broad-band energy coverage and the 
good spectral resolution ($\delta E/E \sim 0.02 (5.9{\rm~keV}/E)^{0.5}$). 
The spatial resolution of the data is, however, limited by the poor 
point spread function (PSF) of the \asca telescope. Although the 
PSF has a sharp core with a FWHM of $\sim
1^{\prime}$, the broad wing of the PSF produces a 50\% encircled radius of
$\sim 1\farcm5$ (Jalota, Gotthelf, \& Zoonematkermani
1992). The PSF also contains some energy dependence which is most
notable above $\sim 6$ keV. The broad wing
of the PSF prevents an {\sl independent} measurement of the background
in the region of the source uncontaminated by the source spectrum
itself. 

	We thus devised a method that simultaneously solves for
both the source and background spectra (see Appendix A for
details). We applied this method to determine the
\asca spectra for both N157B and SNR 0540-69.3.  
To minimize the uncertainties in the spectral background subtraction, 
we chose a small on-source aperture of radius 1\farcm7, and a concentric 
background annulus with its inner and outer radii of 2\farcm3 and 4\farcm9. 
We further used only the southwest half of 
the background annulus around N157B to avoid the possible contamination by 
the 30 Dor nebula in the northeastern half of the annulus. 
Within the on-source aperture of this remnant, the background accounts for 
about 7\% of the total observed intensity with a relatively soft 
spectrum. Judging from the background intensity variation around
N157B, the uncertainty in the background should be less than $\sim 50\%$.
This uncertainty in the background subtraction 
is probably within the systematic error of the data and the analysis. 

For each SIS data set of N157B we generated an appropriate background
spectral file to match the on-source
file. These spectra were then summed for both CCD detectors and for both
observations. Individual response files were summed to create an 
appropriate count weighted multi-CCD detector and multi-telescope 
response matrix. The summation was accomplished using the FTOOL
MATHPHA to add \asca\ spectral files, and the combined
instrument/response files were merged using the FTOOL ADDRMF\footnote{
Information of the FTOOLS software package is 
available in the \asca data analysis guide (Day et al. 1995).}. The
spectra were binned to contain at least 25 counts for $\chi^2$ model
fits. Our final spectral analysis of N157B was based on this
summed spectrum and response. We also conducted fits to the four 
sets of spectra before the summation, and found that the results were 
consistent with those from fits to the summed spectrum, which was
certainly much easier to analyze and made multi-component
model fits possible.

	The GIS data are the most suitable for timing analysis to search for 
a putative pulsar in N157B. All the GIS data, acquired in the highest 
PH time resolution mode (10 bit), sets a limit on the timing accuracy during
space-craft telemetry down-links of ~0.3 milli-seconds for HIGH
bit-rate. However, the calibration of the absolute time (ASCATIME), is
only considered accurate to about 2.0 ms (Hirayama et al.  1996,
\asca news no. 4). A pulsar search which combined high and medium
bit-rate mode data limits the overall accuracy to 5 ms; low-bit rate
mode data was excluded from our pulsar search. We converted the
arrival time of each count to the barycentric dynamical time, using
the FTOOL task TIMECONV. We also extracted
a GIS light-curve from the \asca observation of PSR 0540-69.3 to be
used as a control on our procedure.

	The SIS data are of limited use in timing analysis. The 4 CCD 
mode SIS observation of N157B used a 16 s integration (time
between CCD read-out), setting a fundamental limit on searches for
frequencies faster than 0.03125 Hz, the appropriate Nyquist frequency.

\subsection {\rosat Observations}

	We utilized two \rosat observations from the \rosat High Resolution 
Imager (RHRI) and one from the \rosat Positional Sensitive Proportional 
Counter (PSPC). Both the PSPC and RHRI were sensitive to photons in 
the $\sim 0.1-2$~keV range. 

	The RHRI observations with a PSF of $\sim 6^{\prime\prime}$ 
(FWHM) are the best for studying the spatial properties of N157B. One of the 
RHRI
observations (\rosat Seq. No. rh600228) with an effective exposure of 30094~s 
was pointed at R136, $\sim 7^{\prime}$ away from N157B; the 
other (wh500036) of 4460~s was pointed directly at N157B. Each RHRI 
observation covers a field of $\sim 34^\prime$ diameter. A coadded intensity 
image of the two observations is presented in Fig.~2. The
processing of the observations was described in an earlier article
by Wang (1995), where the same two observations were used to detect 
two point-like X-ray sources in the 30 Dor core region and to identify them
as Wolf-Rayet + black hole binaries. Briefly, small astrometric errors 
($\sim 3^{\prime\prime}$) were corrected through comparisons between optical 
and X-ray positions of identified X-ray sources in 
the region. Therefore, the absolute astrometric accuracy of the
observations is better than $\sim 3^{\prime\prime}$. 

	The PSPC observation (rp500131), targeted at R136 (Chu et al. 1993), 
was used to offer a useful combination of moderate spatial resolution 
(FWHM $\sim 0\farcm5$) and spectral resolving capability 
(about six independent bands in the 0.1-2~keV range). 
The spatial resolution of the PSPC, though not as good as
that of the RHRI, was still significantly better than the \asca
instruments. We extracted a spectrum of N157B from the PSPC 
observation, using an aperture of 1\farcm5 for the on-source 
count rate and an annulus between 2$^\prime$-3\farcm5 radii for the 
background estimation. The aperture is large enough
to contain essentially all the counts from the remnant. 
The total background-subtracted count rate is $0.24 {\rm~counts~s^{-1}}$
in the PSPC 0.1-2~keV band. Similarly, we obtained the RHRI count 
rate of $0.083  {\rm~counts~s^{-1}}$ in the same aperture. 

	We found that the PSPC and RHRI observations provided no
strong constraints on the timing properties of N157B. The observations
were taken in various time segments spreading over a long period of time --- 
about a half year in the case of rh600228. 

\section {Analysis and Results}

\subsection {Timing Properties}

	We conducted an extensive search for pulsed emission from N157B. 
To familiarize ourselves with various methods used, 
we applied them first to SNR 0540-69.3, which contains
a known pulsar. It was easy to detect the pulsar's period 
of 0.05041843~s, even with one tenth of the SNR's data.
Above 2~keV, the pulsed fraction is $26\pm2\%$, almost independent of
photon energy. But the fraction falls to $15\pm2\%$ in the 1-2~keV range
and to $9\pm5\%$ below 1~keV, apparently because of soft diffuse emission 
from the remnant. No significant periodic signal, however,  was detected 
from N157B, confirming the preliminary conclusion reached by Itoh et al. 
(1994). In the following, we concentrate on describing our timing
analysis of N157B with the GIS data. 

We made an FFT analysis of the combined light curve of the two GIS detectors.
This analysis is sensitive to the period range from 3~ms to about 2000~s.
The lower boundary is chiefly due to the number of time bins allowed by 
our computer memory, and the upper limit to various
observing intervals or gaps on timescales comparable to, or greater than,
2000~s. A single FFT of the light curve is the most sensitive, but is
limited to the period range longer than 0.05~s. For shorter periods, we first
divided the total observing time into up to 16 equal time intervals, and then 
applied FFT separately to the individual light curves. The derived
power spectra were averaged to form a combined spectrum, where we 
found no significant peak. We even tried the incoherent harmonics addition 
(e.g., Lyne \& Graham-Smith 1990) to strengthen  
harmonics, in case the pulse shape was narrow. 
But no consistent harmonic pattern was discovered.

	We further utilized a period folding method, but again detected 
no significant signal. A $\chi^2$ fit with a sinusoidal
waveform to a folded pulse profile at an FFT peak period typically yielded
a pulsed fraction of $\sim 6\%$. By re-sampling counts according to the 
``best-fit'' model, we estimated a 99\% upper limit to the pulsed fraction 
as $\sim 9\%$ in the energy band $2-10$~keV. The limit for a sharper
pulse profile would be tighter. A similar analysis with the PSPC data
gave an upper limit as $\sim 17\%$ in the 0.5-2~keV band. 

\subsection {Spatial Properties}

	Figs.~1 and 2 provide global views of the region including both N157B 
and the 30 Doradus nebula. The RHRI image shows that N157B is well 
separated from the main body of the 30 Dor nebula. But
in the SIS map, scattered X-rays can be considerable near N157B 
at energies $\lesssim 1.5$~keV. 

	 Fig.~3 presents close-ups of N157B in X-ray, compared with images 
in H$\alpha$ and in radio continuum. The northeastern
half of the PSPC image (Fig.~3a) shows a clear excess of diffuse X-ray
radiation, but this diffuse enhancement is harder to see in the 
raw distribution of RHRI counts (Fig.~3b), which include a large number
of non-cosmic X-ray events. The low surface brightness diffuse emission is, 
however, evident in the 
smoothed RHRI maps (Figs.~3c-d). The distinct part of the enhancement, outlined
by the lowest solid contour 
level ($7.4 \times 10^{-3} {\rm counts~s^{-1}~arcmin^{-2} }$) in Fig.~3c, 
is somewhat square-shaped, and has a dimension of 
$\sim 85^{\prime\prime}\times 85^{\prime\prime}$. 

	On top of the low surface brightness diffuse X-ray emission is a 
comet-shaped X-ray feature.  The extended
morphology of this feature, similar in the two RHRI observations, 
cannot be due to any instrumental effects. Point-like
objects in the field do not show such elongation (Fig.~2; Wang 1995). 
This feature has a position angle of $\sim 38^\circ$ west to the north. At
the southeastern end of the feature is a compact source 
($R.A. = 5^h 37^m47^s.6; Decl. = -69^\circ 10^\prime 20^{\prime\prime}$),
which appears extended in rh600228. A maximum likelihood fit to the 
count distribution of the source suggests count rate of $(25 \pm 5)
\times 10^{-3} {\rm~counts~s^{-1}}$ and a PSF-subtracted 90\% upper limit 
to the source's full Gaussian size as $\sim 7^{\prime\prime}$. 
But an extent of a few arcseconds could be caused by the pointing 
``jitter'' of \rosat. Thus, the source is only marginally resolved, at 
the best. 

	A different perspective of the data is presented in Fig.~4, where
individual cuts of the RHRI intensity distribution are plotted. 
The compact source of the comet-shaped X-ray feature appears broader than the 
RHRI PSF (Fig.~4a), although part of this broadening may be due to the 
pointing uncertainty in the observation
rh600228. The broad shoulder on the right (northwestern) side of the 
peak gives an estimate of the length of the X-ray feature as $\sim 
30^{\prime\prime} \pm6^{\prime\prime}$. The total width of the feature is 
$\sim 20^{\prime\prime}\pm 4^{\prime\prime}$ (Fig.~4b). Table 2 summarizes 
the properties of the individual X-ray components. 

\subsection {Spectral Properties}

	Figs.~5 and 6 show a remarkable similarity
between the SIS spectra of SNR 0540-69.3 and N157B.  Both spectra are flat
and contain no apparent line features above 1.5~keV, setting these two 
remnants apart from young shell-like LMC SNRs studied by 
Hughes et al. (1995). In the SIS spectra of the shell-like SNRs, strong lines
{\sl dominate}, and weak continua fall steeply above $\sim 1$~keV.

The spectra of SNR 0540-69.3 and N157B show low energy turnovers 
at $\sim 1$~keV, indicating that the absorption is mainly due to oxygen and 
neon. The interstellar abundances of these elements are measured 
typically within the range between 30-40\% 
solar in the LMC (Russell \& Dopita 1992; de Boer et al. 1985), and 
around 40-60\% solar in the solar neighborhood (e.g., Meyer 1995).
Because the \ion{H}{1} column density of the Milky way in the direction
is only $3.2\pm 1.3 \times 10^{20} {\rm~cm^{-2}}$ (Bessell 1991),
the absorption in the LMC dominates.
We assumed an average metal abundance of 40\% solar for X-ray-absorbing gas 
(Morrison \& McCammon 1983) toward both N157B and SNR 0540-69.3. 
A slightly different choice of the metal abundance would only affect
the equivalent hydrogen column density $N_H$ in a spectral fit, but
would hardly affect the fitting of other spectral parameters. 
Table 3 summarizes  spectral parameters from various fits to the 
spectral data of the two remnants. 

The spectrum of SNR 0540-69.3 is well fitted with a simple power law. 
The power law slope is in good agreement with that 
obtained by Finley et al. (1993), whose analysis is based on a PSPC spectrum 
and assumes a solar metal abundance. Switching the abundance in 
our fit to the solar value and fixing the slope to their best-fit value 1.0, 
we obtained almost the same $N_H$ and 0.1-2.4~keV band flux as 
theirs. This comparison suggests no gross inconsistency between the 
\rosat and \asca data. 

	A power law, however, does not fit the spectrum of N157B as 
satisfactorily as that of SNR 0540-69.3. The power law fit of the N157B 
spectrum can be rejected at the $\gtrsim 99\%$ confidence. The
residuals of the fit (Fig.~6) exhibit significant systematic fluctuations 
on energy scales of $\sim 0.1-0.3$~keV, indicating possible emission 
line contributions from an optically-thin thermal plasma. The feature 
at $\sim 0.9$~keV, however, appears a bit too narrow, and its astronomical 
nature may be questioned, although a stronger line of similar width 
at $\sim 1$~keV has been reported in the spectrum of the nearby X-ray 
binary system 4U1626-67 (Angelini et al. 1994). Other possible line 
features are at $\sim 1.4$~keV and $1.8$~keV. 

	To probe the significance of the line features and the
thermal plasma component in general, we 
made a joint spectral analysis of the SIS and PSPC spectra of N157B. 
First, we fitted the two spectra with a power law, allowing the model 
normalizations to vary independently. A significant excess around $0.9$~keV 
is apparent in the PSPC spectrum (Fig.~7),
even greater than that in the SIS spectrum. 
An inclusion of a simple Gaussian line improves the fit 
significantly, reducing the $\chi^2/d.o.f.$ from 197/150 to 171/146; 
the excess is gone in the residuals for both the SIS and PSPC spectra 
(Fig.~8; Table~3). The line in the PSPC spectrum
has a flux 3.5 (2.4-5.5)$\times 10^{-4} {\rm~photons~s^{-1}~cm^{-2}}$ 
at $0.9(0.89-0.92)$~keV. The flux in the SIS spectrum is, however, only 
45(21-68)\%  of that in the PSPC spectrum,
whereas the power law normalizations in the SIS and 
PSPC spectra are very close (within $\sim 5\%$). This line 
flux discrepancy can be explained if the thermal component originates in the 
low surface brightness diffuse emission region (Table~2). 
Part of the diffuse emission is not accounted for in our extraction of 
the SIS spectra, which assumes the remnant to be point-like. 

	We further tried the standard collisionally equilibrium plasma model 
(i.e., Raymond \& Smith 1994, R\&S hereafter). For simplicity, 
we discuss here only the results from the fits to the SIS spectrum. A
single-temperature model fit is not satisfactory; the best-fit 
model is much steeper than the
SIS spectrum at energies $\gtrsim 4$~keV. Such a thermal model is,
however, perfectly acceptable for the 
{\sl Einstein} SSS data, partly because of the instrument's narrower 
energy coverage $\lesssim 4$~keV. The absorption inferred from the SSS 
spectrum is also significantly smaller than the value obtained from the 
\asca data. We suspect that this is caused by a
considerable soft X-ray contribution from the main body of 30 Dor
within the SSS aperture of $\sim 6^\prime$ diameter. Therefore, the
results from the SSS spectrum should be used with caution.
A two-component model consisting of
a power law and an R\&S plasma improves our power law fit to the SIS spectrum
only slightly ($\chi^2/d.o.f. = 164/125$); the thermal component, 
even with the abundance allowed to vary, does not 
adequately account for the apparent line features. Without accurate 
information on the remnant's environment, structure, and evolution,
we are unable to use the limited statistics and line signatures of the data 
to explore the vast parameter space of more sophisticated models. 

	The power-law contribution may be determined, however, if it
is solely due to the comet-shaped feature, presumably a synchrotron nebula 
(\S 4.3). We used the RHRI count rate of $0.054 {\rm~counts~s^{-1}}$ (Table~2; 
with an estimated uncertainty of $\sim 20\%$) to isolate the contribution of 
the feature below 2~keV. At higher energies, where the thermal contribution 
may be neglected, we used the SIS spectrum. The spectral parameters from 
a power fit to this SIS/RHRI combined spectral data are
included in Table 3. Because of the 
exclusion of the diffuse component, the $N_H$ value is
greater than those from the previous fits. The luminosity of the power 
law component is $L_x = 1.0 \times 10^{36} {\rm~ergs~s^{-1}}$ in the 
2-7~keV band, or $L_x = 5.6 \times 10^{36} {\rm~ergs~s^{-1}}$ in the 
0.2-4~keV band. The uncertainties of  these luminosity estimates are
$\sim 30\%$.

Fig.~9 presents the power law model, together with the SIS spectrum including 
data points below 2~keV. The residuals, evidently showing line features, 
do appear to be thermal in origin. A fit with a single temperature R\&S model
suggests a characteristic plasma temperature of $\sim 0.7$~keV. However,
the fit ($\chi^2/d.o.f. = 187/128$) is not satisfactory; there is a
significant excess above the model below 0.7~keV.
A similar fit ($\chi^2/d.o.f. = 21/26$)
to the residuals in the PSPC spectrum is good, and the best-fit temperature is
$0.41 (0.31-0.53)$~keV. Thus, the thermal model describes the residuals
well only at the PSPC's resolution. The predicted RHRI count rate 
$\sim 0.031 {\rm~counts~s^{-1}}$ is consistent with the rate of the diffuse 
component within an estimated systematic error of up to $\sim 40\%$.

	The above isolation of the power law component is physically plausible,
but not unique. Alternatively, the power law component might arise only in 
the compact X-ray source. A power law fit to this combination (Table 3), 
however, is not as satisfactory. In particular, the absorption
appears much too large (by a factor of $\sim 3$) 
to be consistent with the  measurement of the optical 
extinction toward N157B (\S 4.4). 

\section {Discussion}

\subsection{Qualitative Considerations}

The striking similarity between the spectra of N157B and SNR 0540-69.3,
together with the centrally-peaked morphology of N157B in both radio and X-ray,
leaves little doubt about N157B as a Crab-like SNR.
In contrast to the highly polarized emission observed in the Crab and 
SNR 0540-69.3, however, no radio polarization has been detected in 
N157B (D94). 

	The comet-shaped X-ray feature is evidently the most interesting 
emission component of N157B. Markwardt \& \"Ogelman (1995) have
observed a morphologically similar feature near 
the Vela pulsar. They interpret this feature as
the thermal emission from a cocoon around a jet from the pulsar. 
The pulsar's proper motion is found to be nearly perpendicular to the 
direction of the jet. Therefore, it is possible that the jet is sweeping up 
the ambient material, which may explain the apparent higher density in 
the cocoon than in the ambient medium. However, the X-ray 
luminosity of the N157B feature is more 
than three orders of magnitude greater than that 
($\sim 10^{33} {\rm~ergs~s^{-1}}$) of the Vela jet. Furthermore, 
the {\sl ASCA} spectrum shows that the emission, most of which arises
in the comet-shaped feature, is predominantly nonthermal in origin (\S 3.3). 

\subsection {Low Surface Brightness Diffuse X-ray Emission}

	The diffuse emission component of N157B most 
likely arises in a diffuse plasma, which is responsible for the possible weak 
line features in the \asca spectrum. We adopt a conversion of an RHRI count rate 
to an emission integral (EI --- defined as $\int n_e^2 {\rm~d}V$)
as $2.7 \times 10^{60} {\rm~cm^{-3}/(counts~s^{-1})}$.
This conversion applies to the model of $N_H \sim 10^{22} 
{\rm~cm^{-2}}$ and $ T \sim 0.7 {\rm~keV}$, and is good within 
a factor of 2 as long as the temperature $\gtrsim 0.4$~keV. 
From Table 2, we infer a count rate of the diffuse component as $0.019 
{\rm~counts~s^{-1}}$, corresponding to an EI of $5.0 \times 10^{58} 
{\rm~cm^{-3}}$ and an X-ray luminosity as 
$1\times 10^{36} {\rm~ergs~s^{-1}}$ in 0.1-2.4~keV band, or $0.7
\times 10^{36} {\rm~ergs~s^{-1}}$ in the 0.5-2~keV band.
Further assuming that the diffuse component arises in an approximately 
spherical space of diameter $D = 20$~pc, and using $\zeta$ as the 
effective spatial filling factor, we estimate
the gas density $n_h$ as $\sim 0.6 \zeta^{-0.5}
{\rm~cm^{-3}}$.  The corresponding total gas mass is then $\sim 56 \zeta^{0.5}
{\rm~M_\odot}$. If $\zeta \gtrsim 0.1$, the remnant expansion should then be
in the Sedov phase (e.g., Cox 1972). The total thermal energy in the 
X-ray-emitting plasma is $\sim 3 \times 10^{50} \zeta^{0.5} 
{\rm~ergs}$, a considerable fraction of the expected mechanical energy of an 
SN. Because this fraction is expected to be $\sim 72\%$ in the 
Sedov phase, the inferred SN energy is in a
nominal range of a few times $10^{50} {\rm~ergs}$. Finally, the 
Sedov age of the remnant is $t_{SNR} = 5 \times 10^3 D_{20 {\rm pc}} T_{0.7
{\rm keV}}^{-1/2}$~yrs, where $D_{20 {\rm pc}}$ is the remnant diameter
in units of 20~pc and $T_{0.7
{\rm keV}}$ the diffuse plasma temperature in units of 0.7~keV.

Both the above X-ray size and boundary agree 
with those inferred from kinematic studies of H$\alpha$-emitting gas 
in N157B (e.g., CKSL92). These studies have revealed shocked materials 
with velocity offsets up to $\sim 10^2 {\rm~km~s^{-1}}$.
However, since the velocity of H$\alpha$-emitting gas does not necessarily 
represent the expansion of the outer shock of the N157B remnant, the inferred 
characteristic age can, at most, be considered as an upper 
limit to the true age of the remnant. 

\subsection {Comet-Shaped X-ray Feature}

	The predominant non-thermal nature of N157B leads
us to explain the comet-shaped X-ray feature as a
synchrotron nebula, powered by the putative pulsar embedded in N157B. 
This explanation, constrained by the observations, provides 
specific predictions about the energetics and dynamics as well as the 
spatial and spectral properties of the remnant.

From the luminosity of the X-ray feature, 
$L_x = 5.6 \times 10^{36} {\rm~ergs~s^{-1}}$ in the 0.2-4~keV band (\S 3.3),
we may estimate the spin-down power $\dot{E}$ of the pulsar. A linear
relation ${\rm log}L_x \approx 1.39~{\rm log}\dot{E}-16.6$, obtained by
Seward \& Wang (1988), fits reasonably well to the three confirmed 
Crab-like SNRs. Applying this relation to N157B, we obtain $\dot{E} \approx
2 \times 10^{38} {\rm~ergs~s^{-1}}$. Following Seward \& Wang (1988) again,
we further estimate the period and magnetic field strength of the pulsar
as  $\sim (0.08 {\rm~s}) (t_3 \dot{E}_{38})^{-1/2}$ and $\sim (1 
\times 10^{13} {\rm~G}) t_3^{-1} \dot{E}_{38}^{-1/2}$, where $t_3$ is the 
remnant's age in units of $10^3$~yrs. 

	The steep X-ray spectrum provides clues about the energy losing
mechanism of relativistic particles in the synchrotron
nebula. The X-ray-emitting lifetime $t_x$ of an electron in the nebula 
can be expressed as 
\begin{equation}t_x \sim (40 {\rm~yrs}) 
\epsilon^{-1/2} (H_{-4})^{-3/2},\end{equation}
where $\epsilon$ is the characteristic synchrotron photon energy in units of 
keV, and $H_{-4}$ is the magnetic field in units of $10^{-4}$~G. This timescale is 
much shorter than the expected age of the remnant. Therefore, a steady 
pulsar spin-down luminosity is a good approximation to model
the current X-ray emission from the nebula. We assume that the distribution
of pulsar injected particles is approximately a power law with an index of
$\gamma$, and that the velocity distribution is isotropic.
The X-ray emission spectrum is then a power law with $\alpha = \gamma/2$ if 
synchrotron radiation loss dominates, or $\alpha = (\gamma-1)/2$ if the 
energy loss is predominately due to
adiabatic expansion of pulsar wind materials (e.g., Tucker 1977). 
In general, one expects a convex-shaped
spectrum: The adiabatic loss dominates at low energies, whereas the 
synchrotron becomes important at high energies. Such a slope change
from high to low energies is a general characteristic of 
Crab-like SNRs (e.g., Du Plessis et al. 1995).
The steep spectrum observed in N157B thus suggests that the synchrotron loss 
dominates in N157B even in the relatively low energy range 0.5-7~keV. 
From the slope of $\sim 1.5$ (\S 3.3), we obtain $\gamma \sim 3$.

	The spatial properties of the comet-shaped X-ray feature have
implications for the dynamics of relativistic electrons. Because of
the short lifetime of the electrons, the size of a synchrotron nebula 
should typically be much smaller than that of the corresponding radio one if
electrons diffuse through the nebula medium at the Alfv\'en velocity. 
The observed size and comet-like shape of the X-ray feature
suggest that X-ray-emitting particles are undergoing an outflow, which can 
travel much faster than diffusing electrons. The compact source, 
where energy is likely being deposited by the pulsar, is separated from 
the radio peak (\S 4.4), where most of the pulsar
spin-down energy has supposedly been dumped. 
If the {\sl birth} place of the pulsar is close to the position of the radio 
peak and has an age of $\sim 5 \times 10^3 {\rm~yrs}$ (\S 4.2),
the separation of $\sim 12^{\prime\prime}$ between the pulsar and 
the radio peak then implies that the pulsar has a transverse velocity of
$\sim 6 \times 10^2 {\rm~km~s^{-1}}$, compared to an average pulsar 
velocity of $\sim 4 \times 10^2 {\rm~km~s^{-1}}$ (Lyne \& Lorimer 1994). 
We use $v_{p,3}$ (in units of $10^3 {\rm~km~s^{-1}}$) to characterize
the total velocity of the pulsar motion. 

Such a relatively fast moving young pulsar can produce a strongly elongated 
pulsar wind bubble (Fig. 10). This bubble consists of several distinct
regions. First, one may expect a bow shock running 
ahead of the pulsar, while the corresponding reverse shock terminates 
the free pulsar wind. The size of the region enclosed by this bow shock 
can be small, but the pressure can be large, depending on the 
ram pressure. Enhanced synchrotron radiation can then make the region 
especially bright in X-ray. Second, accelerated by
the pressure gradient between the bow shock and the ambient medium, the 
shocked wind materials, which are relativistically hot, naturally
forms a supersonic jet inside a tunnel in the opposite direction of the
pulsar's motion (Wang, Li, \& Begelman 1993). Third, the jet
likely shoots through the bubble, and then induces another
bow shock (or a series of oblique shocks)
on the other side. Fourth, the terminated wind materials are forced back to 
the bubble, where most of the pulsar spin-down luminosity is dumped. 
Let us now examine this picture in a more quantitative way.

	The size of the bow shock region around the pulsar should be
consistent with the observed upper limit. 
The  characteristic reverse shock radius $r_s$ is defined by 
the ram-pressure balance between the pulsar wind and the
pulsar motion, 
\begin{equation}r_s = (4 \times 10^{-2} {\rm~pc}) 
[\dot{E}_{38}/(n_a v_{p,3}^2)]^{1/2},\end{equation}
where $\dot{E}_{38}$ is the spin-down luminosity in units of 
$10^{38} {\rm~ergs~s^{-1}}$ and $n_a$ is 
the ambient medium density, which may be of the same order of magnitude
as $n_h$ estimated in the diffuse X-ray emission (\S 4.2). The size 
and shape of the outer bow shock depend
critically on the dynamical effect of magnetic
fields, and no theoretical model has yet been developed to describe
this phenomenon. We thus characterize the total volume of the region
enclosed by the bow shock as $V_x = {4 \over 3} \pi (\xi r_s)^3$, where
the parameter $\xi$ is to be constrained. Assuming that 
the energy is partitioned equally between
particles and magnetic fields, that the field is randomly distributed,
and that the pressure is balanced in the region, we have 
\begin{equation}H \approx (0.8 {\rm~mG}) 
(n_a v_{p,3}^2)^{1/2}\end{equation}
and
\begin{equation}W_e = {H^2 \over 8\pi} V_x, \end{equation} 
where $W_e$ is the total energy in particles.
Since only about half of the observed X-rays from the feature originates
in the compact X-ray source (Table~2), the time
spent by the shocked pulsar wind materials in the bow shock region is probably
comparable to the lifetime of
an electron that emits 1~keV photons. For a steady jet, we expect
\begin{equation}W_e \sim t_x \dot{E} = (5 \times 10^{44} {\rm~ergs}) 
\dot{E}_{38} (n_a v_{p,3}^2)^{-3/4}\end{equation}
From  (1)-(5), we obtain 
\begin{equation}\xi \sim 3 \dot{E}_{38}^{-1/6} (n_a v_{p,3}^2)^{-1/12} 
\end{equation}
The bow shock has a size $\sim 2\xi r_s \sim 1^{\prime\prime} 
\dot{E}_{38}^{1/3} (n_a v_{p,3}^2)^{-7/12}$, consistent with 
the upper limit to the full size of the compact X-ray source 
($\lesssim 7\arcsec$; Table 2).

	After being forced out from the bow shock, the wind materials can 
continue to radiate in X-ray (Fig.~10). The radiation
from the relativistic jet itself (Wang, Li, \& Begelman 1993), 
however, may not be significant. Most of the X-rays from the comet-shaped feature
outside the compact X-ray source are probably released after the jet is 
terminated and are from 
shocked wind materials inside the bubble, explaining why the 
width of the X-ray feature (Table 2) is comparable to that of the compact
radio emission region (\S 4.4). Just like particles in the bow shock region,
the shock materials in the wind bubble are still relativistically hot, and the
emission is primarily synchrotron.

\subsection {Radio Emission}

	We speculate that the lack of polarization in the observed radio 
emission of N157B is due to the presence of a giant \ion{H}{2} region produced 
by the coexisting OB association LH99 (Schild \& Testor 1992). CKSL92 find that
about 20\% of the radio emission arises in the \ion{H}{2} region, to be
consistent with both the H$\alpha$-to-radio continuum ratio and the average
extinction ($A_v = 1.0$~mag), inferred from Balmer decrement measurements 
over a 4\farcm9 region surrounding N157B (Caplan \& Deharveng 1985). The
rest of the radio emission is presumably nonthermal; part of the emission
should be associated with the synchrotron nebula of the N157B remnant. 
The radio emission above the level of $28 {\rm~mJy~beam^{-1}}$ in Fig. 3d 
has an extent similar to the X-ray 
feature, but the centroid of the emission feature is about 
12$^{\prime\prime}\pm3^{\prime\prime}$ away from the compact X-ray source.
Based on the relations $ A_v/E(B-V) = 3.1$ and 
$N_H/ E(B-V)= 2.4 \times 10^{22} {\rm~cm^{-2}}$, which are appropriate
for the interstellar gas in the 30 Dor region (Fitzpatrick 1986), 
we infer $N_H = 7.7 \times 10^{21} {\rm~cm^{-2}}$.
In comparison, our X-ray-measured $N_H$ is  
$\sim 1 \times 10^{22} {\rm~cm^{-2}}$ (\S 3.3). Thus 
the N157B remnant is likely located in or behind the \ion{H}{2} region.

	The ionized gas and the expected magnetic field of the \ion{H}{2} 
region can cause Faraday rotation in the radiation from the 
synchrotron nebula. Assuming that the 20\% of the observed radio emission
arises in a spherical region of $ \sim 20$~pc diameter, similar to
the projected dimension, we estimate a mean free
electron density $n_e \sim 50 {\rm~cm^{-3}}$.
The Faraday rotation angle varies with  $n_e H_{\|}$,
where $H_{\|}$ is the line-of-sight intensity of the field. 
Following Burn (1966), we express the dispersion of the rotation 
within a radio survey beam as
\begin{equation}\delta\phi \sim \left[K n_e H_{\|} d_e 
\lambda^2\right] d_s/d_e,\end{equation}
where the coefficient $K = 2.6 \times 10^{-17}$, $d_e$ is the characteristic
size of turbulent eddies, $d_s=1.8$~pc and $\lambda = 13$~cm are the beam size 
and the wavelength of the 2.3~GHz survey. In the bracket of the above
equation is the Faraday rotation angle through a single eddy. 
Assuming a typical interstellar field of 
$H_{\|} \sim 5 \mu$G, we obtain $\delta \phi \sim 6$. This dispersion 
produces a radiation depolarization 
\begin{equation} p \sim p_i \ exp[-2(\delta\phi)^2 d_{HII}/d_e],\end{equation}
where $d_e$ should be comparable to, or smaller than, the dimension $d_{HII}$
of the region. Therefore, only if a fraction of 
the observed \ion{H}{2} region is in front
of the nebula can the Faraday dispersion explain the lack of
the polarized emission in the 2.3~GHz survey. Clearly, a radio 
survey at a higher
frequency will have a better chance to detect the polarized emission from
the synchrotron nebula.

\section {Summary}

	Based on the \rosat and \asca\ observations, we have examined the 
X-ray timing, spatial, and spectral properties of N157B. An algorithm 
has been devised for extracting \asca\ spectra for a point-like X-ray source.
For comparison, we have also presented an \asca\ spectrum of SNR 0540-69.3, 
the only LMC SNR that contains a detected pulsar. We have arrived
the following main results and conclusions.

\begin {enumerate}

\item The flat and featureless spectrum of N157B in the $2-7$~keV range
resembles that of SNR 0540-69.3, and is distinctly
different from those of young shell-like SNRs in the LMC.
The power law energy slope ($\sim 1.5$) of the spectrum
is, however, steeper than those ($\sim 1.0$)
of SNR 0540-69.3, MSH 15-52, and the Crab nebula.

\item The \asca spectrum of N157B shows possible line features at 
energies $\lesssim 2$~keV. A joint analysis of the \asca and \rosat 
data suggests that these features are likely associated with a 
thermal emission component of the remnant. 

\item No pulsed signal is identified; 
the pulsed fraction is $\lesssim 17\%$ of the remnant's total luminosity
in the 0.5-2~keV band, or $\lesssim 9\%$ in the 2-10~keV band. So
the putative pulsar in N157B is still elusive.

\item More than two thirds of the observed X-ray radiation from N157B 
comes from a comet-shaped X-ray feature, which has a size of $\sim 7{\rm~pc} 
\times 5{\rm~pc}$ and a luminosity of $\sim 1 \times 10^{36} 
{\rm~ergs~s^{-1}}$ in the 2-7~keV range. 

\item We interpret the X-ray feature as the synchrotron
radiation from pulsar wind materials. The pulsar is likely
moving at a speed of $\sim 10^3 {\rm~km~s^{-1}}$. A ram-pressure confined
bow shock around the pulsar may account for an embedded compact X-ray source,
which is probably unresolved. The ram pressure further focuses fresh wind 
materials into a relativistic jet. The jet is eventually terminated on the 
other side of the pulsar wind bubble, where most of the spin-down energy 
has been dumped. 

\item Around the X-ray feature, we further detect a low surface brightness
diffuse X-ray emission region.
This region most likely accounts for the  apparent thermal component
in the \rosat and \asca spectra below $\sim 2$~keV. 
The remnant has a diameter of $\sim 20$~pc and a 0.5-2~keV luminosity 
of $\sim 7 \times 10^{35} {\rm~ergs~s^{-1}}$, a total thermal energy
of $\sim 3 \times 10^{50}$~ergs,
and an age of $\sim 5 \times 10^3$~yrs.

\item The lack of polarized radio emission from N157B may be explained
by the depolarization of a foreground \ion{H}{2} region.

\end {enumerate}

\acknowledgements

	We thank J. Dickel for the radio map, C. Smith for the H$\alpha$
image used in the paper, and G. Testor for sending us his
optical CCD images of the N157B region. We thank the referees for various 
critical comments that led to improvements in the paper. This work is 
funded by a Lindheimer Fellowship and by NASA grant NAG5-2717. 

\appendix

\section{\asca\ Background Spectral Estimation for an Isolated Point Source}

\subsection{Introduction}

	Estimating a local background spectrum and subtracting it from
an on-source X-ray spectrum normally represents a crucial step in
analyzing the spectral properties of a source.  Because of the broad
\asca point spread function (PSF) (see \S 2.1), it is not usually
practical to obtain an independent measurement of the background free
of source contamination without going far from the source, where data
might not exist, or the background might differ significantly.  For
example, an annular background region of radius 3$^{\prime}$ -
5$^{\prime}$ would require that the source count rate be less than
$0.03$ counts per second for it not to contribute more than 10\% to the
total counts in the background region. This count rate is barely
detectable in a typical 20 ks observation. For an observation of a
moderately bright point source, the counts in the background are an
admixture of source and background counts. On the other extreme, a  source 
of a few count per sec would dominate the background region, contributing
$> 90\%$, and the background can be ignored to first-order. We are
interested here in the intermediate case.

The measured background includes X-rays of cosmic
origin, X-rays and particles of Solar origin, cosmic ray induced X-rays, and
internal instrumental counts. Perhaps most insidious for \asca is the
stray light X-rays (mirror rays which undergoes single, instead of
double reflections, etc.), which are not imaged in the usual way,
and which depends strongly on the celestial distribution of X-ray
emission in and out ($\lesssim 1^{\circ}$) of the telescope's field of
view. The background
depends on the satellite geographical location and height above the
Earth, solar activity, and the incident cosmic X-rays as well as
telescope pointing angle relative to the Earth, Sun, and Earth's
magnetic field vector. Taken
together, these produce a background for a given observation whose
spectrum and intensity depends on date, orbit, sky, and detector
coordinates. The broad PSF wings also mean that other sources in the
field may substantially color the target spectra, producing an
additional effective background.

The current recommended practice for estimating the background is to
extract counts from the \asca Deep Field pointings, screened in a
manner identical to the data, and selected from the source detector
region. For the simple case of an isolated point source, using
the Deep Fields should give a reasonable zero-th order estimation of
the background. However, for a given observation and target, this
practice is limited by the uniqueness of background as discussed
above. As an added complication, although these fields might represent
the background statistically in an average field, they are known 
to contain explicit point
sources, which may compromise extracted spectra near their
detector locations. Furthermore, the response of the detector
to the background changes over the course of the mission due to known
secular degradation of the instruments with time. Most importantly,
the background intensity and spectrum can vary from one field to another,
especially in regions of nearby galaxies such as the LMC.

Here, we present a simple method for a first-order estimate of both
the source spectrum and the background spectrum from the data itself, for the case
of an isolated point \asca source embedded in a statistically uniform
local background. The advantage of self-calibrating is evident given
the strong temporal, instrumental, orbital, and celestial dependence of
the background.  Our method is to consider the spectral ratio of
observed and expected counts in concentric annuli centered on the
source, each of which is an admixture of the true source and
background spectrum. We look for deviations of the radial average
profile from that expected for a point source and simultaneously solve
for the two spatially distinct spectral components. Thus we can
estimate the background over the source region itself. This is in
effect a form of spectro-spatial deconvolution of the background
and source spectra.

\subsection{Method}

Consider the observed counts in two concentric regions centered on the
source and in the measured spectral bin (``PHA'' bin) $E^{\prime}$,
$O_1(E^{\prime})$ and $O_2(E^{\prime})$, both of which are a
composition of the source and background spectra, $S(E^{\prime})$ and
$B(E^{\prime})$ respectively,

\begin{equation}
O_1(E^{\prime}) = \rho_1(E) S(E^{\prime}) + \alpha_1 B(E^{\prime})
\end{equation} 
and,
\begin{equation}
O_2(E^{\prime}) = \rho_2(E) S(E^{\prime}) + \alpha_2 B(E^{\prime}) v(E)
\end{equation} 

where $\alpha_1, \alpha_2 $ are the geometric areas of the respective
regions; $v(E)$ is the mean vignetting of the background 
region relative to the source; $\rho_1(E)$ and
$\rho_2(E)$ are the encircled energy function\footnote{The 
unfortunate use of the term ``energy'' here is not to be
confused with the spectral energy, as it in fact refers to the power
contained in the mirror PSF at
various radii; it is defined as the relative counts (``energy'')
enclosed in the PSF between the radii a and b, $EEF(E;a,b) =
{\int_{a}^{b} PSF(E)} / {\int_{0}^{\infty} PSF(E)}$.}
v(EEFs) within those areas. Here, the energy, $E$, refers to the energy
dependence of $\rho(E)$ and $v(E)$, which also depend on specific
locations of $\alpha_1$ and $\alpha_2$ in detector coordinates. These 
dependencies have been modeled with laboratory and on-orbit calibrations 
(see \S A1.3, below) and therefore, the ratios $\gamma = \alpha_1 / 
v(E)\alpha_2$, and $\delta = \rho_1 / \rho_2$, are determined. Notice 
that these ratio is not strongly energy dependent below $ \sim 6$ keV, 
and for most cases, to the limit of our approximation, a single number 
for each is reasonable.

We can then solve for the source and background spectra, 
and find,
\begin{equation}
\rho_1(E) S(E^{\prime}) = \delta ( O_1(E^{\prime})-\gamma O_2(E^{\prime}) ) / (\delta-\gamma)
\end{equation} 
\begin{equation}
\alpha_1 B(E^{\prime}) = \gamma ( O_1(E^{\prime})-\delta O_2(E^{\prime}) ) / (\gamma-\delta)
\end{equation} 
where $\rho_1 S(E^{\prime})$ is the total counts enclosed in the source area
$\alpha_1$ in spectral bin $E^{\prime}$, and $\alpha_1 B(E^{\prime})$ is the background spectra
over the same area at that energy. The coefficients of $S(E^{\prime})$ and
$B(E^{\prime})$ insure correct normalization and units.  The resulting errors
for the source and background are simply found from the quadratic sum
of the errors associated with individual terms.

In practice, we carry out this calculation on each PHA (PI) bin in our
spectral file. The energy dependencies ($E$) and ($E^{\prime}$) are
related by the spectral redistribution file, RMF($E$;$E^{\prime}$),
and do not enjoy a one-to-one correspondence. However, our method is
insensitive to this redistribution, as it is a third-order effect
because it arises in the errors in the ratios $\gamma$ and $\delta$,
which come from the fraction of redistributed counts at the lower
energies. The typical negative sloped spectra and the weak energy
dependence of these ratios insure that this effect is negligible for
our purpose.

For simplicity, we assume the ratios
$\gamma$ and $\delta$ are energy independent. We can then easily
compute the source and background spectral components using the FTOOL
MATHPHA to perform the calculation on each element of the spectral
(``PHA'') file. We are then left with a spectral file which can be
imported directly into a spectral analysis package such as
XSPEC. These spectra will be
correctly normalized and the instrument and mirror response files
(``RMF'' + ``ARF'')\footnote{A description of the ``PHA'', ``ARF'',
\& ``RMF'' files, the
XSPEC spectral analysis program, and the FTOOLS software package is
available in the \asca data analysis guide (Day et al. 1995)
and references therein. }
created for $O_1$($E^{\prime}$) are still appropriate.  It is
most practical to use the observed source spectra $O_1$($E^{\prime}$) along with
the derived background spectra $\alpha_1 B_1$($E^{\prime}$) in analyzing the
source spectra. 

The energy dependence of $\rho(E)$ and $v(E)$ can be explicitly accounted
for by using the Awaki's ``XRTEA'' effective area subroutine which returns
$\rho(E)$ and $v(E)$ as a function of source position in a detector 
and off-source angle. This code is available in the FTOOLS software package.
An attractive alternative for measuring the EEF is to use an observation of
a bright point-like source. If this fiducial source is acquired at the
same detector position (i.e. 1 CCD mode default) as the source under
study, and the data screened and exposure-corrected in the same way,
then the background for the fiducial source relative to the study source
may be reasonably ignored. Modeling the EEF with a real observation
allows for the jitter in the attitude, which is similar from
observation to observation (order $\lesssim 10^{\prime\prime}$). For the
SIS, the fiducial source should not be too bright as to be noticeably
affected by pile-up. The bright CVs make excellent targets, and have
been successfully used for this purpose.

In conclusion, the above method allows an
estimate of the background spectrum under a circular aperture, centered on a
point-like source. Given enough counts, this method can naturally extend to a
least square solution for a general spectro-spatial deconvolution for
a point source.  We may extract counts from a set of $N$ concentric
annuli and solve the set of linear equations of the form:
\begin{equation}
O_n(E^{\prime}) = \rho_n(E) S(E^{\prime}) + \alpha_n B(E^{\prime}) v_n(E)
\end{equation} 

\begin{table}[htb]
\begin{tabular}{lccccc} 
\multicolumn{6}{c}{\bf Table 1:  Summary of \asca Exposures on  N157B }\\ [0.1in]
\hline \hline
 &    Detector       & Exposure  &  Source   & Off-axis  &  Vignetting\\
 &    (CCD Chip)     &  Time     & Counts\tablenotemark{a}
& Distance  &   Fraction\\			     
 &                   & (s)    &         & (arcmin)  &  \\
\hline   	
\multicolumn{6}{c}{\it \asca Seq. No. 20000000}\\	    	         	     	       		   
 &    SIS0 (C2)      &  16743    &  2961  &  3.06  &    0.90\\
 &    SIS1 (C0)      &  19245    &  2717  &  7.17  &    0.64\\
 &    GIS2           &  30820    &  6784  &  3.43  &    0.87\\
 &    GIS3           &  30816    &  5856  &  7.14  &    0.64\\
\multicolumn{6}{c}{\it \asca Seq. No. 50002000}\\                 	     	       		   
 &    SIS0 (C0+C3)   &  26123    &  2758  &  8.97  &    0.54\\
 &    SIS1 (C2+C1)   &  16876    &  2623  &  4.01  &    0.84\\
 &    GIS2           &  30543    &  5727  &  7.24  &    0.64\\
 &    GIS3           &  30538    &  6171  &  7.12  &    0.64\\
\hline
\end{tabular}
\tablenotetext{a}{ SIS and GIS counts are obtained 
within 1\farcm7 and 4\farcm0 radius circles centered on N157B.}
\end{table}

\begin{table}[htb]
\begin{tabular}{lccl} 
\multicolumn{4}{c}{\bf Table 2: Spatial Components  of 
N157B}\\ [0.1in]
\hline \hline
Component & X-ray Full Size & Count Rate\tablenotemark{a} & Comment\\
\hline
Whole Remnant & $85^{\prime\prime} \times 85^{\prime\prime}$ & 73& $\gtrsim 
7.4 \times 10^{-3} {\rm counts~s^{-1}~arcmin^{-2} }$\\
Comet-Shaped Feature & 30$^{\prime\prime} \times 20^{\prime\prime}$ & 54 & $\gtrsim 
2.2 \times 10^{-2} {\rm counts~s^{-1}~arcmin^{-2} }$\\
Compact source &  $\lesssim 7^{\prime\prime}$ & 25 &Probably unresolved  \\
\hline
\end{tabular}
\tablenotetext{a}{in units of $10^{-3} {\rm~counts~s^{-1}}$.}
\end{table}
\scriptsize
\begin{table}[htb]
\begin{tabular}{llcccc} 
\multicolumn{6}{c}{\bf Table 3: Model Fit to the \asca and \rosat Spectral data\tablenotemark{a}}\\ [0.1in]
\hline \hline
Data 	& 	Model & $\chi^2/d.o.f.$ &Energy slope & Column Density & Norm\tablenotemark{b} \\
\hline
\multicolumn{6}{c}{\it SNR 0540-69.3}\\
SIS	  	&Power law    	 &172/160& 1.04(0.98-1.10)& 7.9(7.4-8.4) &11(10-12)\\
\multicolumn{6}{c}{\it N157B}\\
SIS+PSPC  	&Power law+line &171/146& 1.4(1.3-1.5)& 6.9(6.3-7.5)&3.2(3.0-3.4)\tablenotemark{c}\\
SIS+RHRI(feature)&Power law 	 &82/78	 & 1.5(1.4-1.7)& 11(7.3-16)  &3.5(2.8-4.5)\\
SIS+RHRI(source)  &Power law 	 &92/78	 & 1.9(1.7-2.1)& 28(22-35)   &6.1(4.8-7.4)\\
\hline 
\end{tabular}
\tablenotetext{a}{The parameter limits  are all at the 90\% confidence level.} 
\tablenotetext{b}{At 1~keV and 
in units of $10^{-3} {\rm~photons~s^{-1}~keV^{-1}~cm^{-2}}$.}
\tablenotetext{c}{From the normalization of the PSPC spectrum.}
\end{table}

\clearpage
\begin{figure} \caption{\asca SIS maps of the 30 Dor region.
The maps are corrected for exposure and vignetting. The contours are at 
(3.0, 4.0, 5.5, 7.8, 11, 16, 24, 35, 52, 77, and 116) $\times 10^{-3}$
${\rm~counts~s^{-1}~arcmin^{-2}}$ in the $0.4
- 1.5$ keV band, and at 1.7, 2.3, 3.2, 4.6, 6.6, 9.6,
14, 21, 31, 47, 69, 104, and 156) $\times 10^{-3}$
${\rm~counts~s^{-1}~arcmin^{-2}}$ in the $1.5-7$ keV
band. Faint radial strips near the southeastern boundaries 
of the map, especially in the $1.5-7$ keV band, are due to stray light 
(reflected mirror rays) from LMC X-1 to the south.
\label{fig2}}
\end{figure}

\begin{figure} \caption{N157B and its environment: RHRI
X-ray intensity contours overlaid on an H$\alpha$ image. The X-ray 
intensity is calculated by coadding the two background-subtracted
RHRI observations and is
adaptively smoothed with a Gaussian of adjustable size to achieve 
a constant local count-to-noise ratio of $\sim$ 8 over the image.
The X-ray contours are at (2.1, 3.8, 5.9, 8.4, 12, 16, 22, 30, 60, 120,
250, 450, and 800) $\times 10^{-3} {\rm~counts~s^{-1}~arcmin^{-2}}$.
\label{fig1}}
\end{figure}

\begin{figure} \caption{Close-up views of the N157B 
region: the PSPC count distribution in the 0.5-2~keV 
band (a), the RHRI count distribution (b), the RHRI 
intensity contours compared with an H$\alpha$ gray-scale map (c), and 
2.3~GHz continuum intensity contours overlaid on the gray-scaled RHRI map 
(d). In (c), the X-ray intensity background-subtracted and
exposure-corrected, is smoothed with a count-to-noise ratio of $\sim 4$,
and the contours are at (3.6, 7.4, 13, 22, 35, 54, 120, 250, 450, and 800)
$\times 10^{-3} {\rm counts~s^{-1}~arcmin^{-2} }$. In (d),
the gray-scaled X-ray intensity is plotted logarithmically 
in the range between $ 2.4\times 10^{-3}$ (about $1\sigma$ above the local
background) and 0.5 
${\rm counts~s^{-1}~arcmin^{-2} }$, and the 2.3 GHz radio continuum contours 
are at 2, 4, 8, 16, 28, 44, 64, and 88 mJy~beam$^{-1}$. 
The radio data (see also D94) have a FWHM beam size of $\sim 7\farcs5$, 
comparable to that of the RHRI image. 
\label{fig3}}
\end{figure}

\begin{figure} \caption{RHRI X-ray intensity distribution in strips across
the field of N157B. Each strip is $15^{\prime\prime}$ wide. 
The histograms are calculated so that each bin has at least 25 
counts. Fig.~4a presents a single SE-NW cut through the feature 
($38^\circ$ west to the north), 
together with the RHRI PSF (the Gaussian-like profile) and the radio 
intensity distribution (solid curve) 
in the same cut; the southeast is to
the left. The label of the ordinate is for the X-ray data, whereas the
radio intensity is in units of ${\rm Jy~beam}^{-1}$).
Fig.~4b includes the NE-SW oriented strips, perpendicular
to the N157B feature. The northeast is to the left, and the most 
northwestern strip is at
the top. The fourth strip is centered on the position of the compact X-ray 
source.
The bottom histogram is in units as posted; for ease of separation,
the intensity in the subsequent histogram is multiplied by 10, and so on. 
\label{fig4}}
\end{figure}

\begin{figure} \caption{\asca SIS spectrum of SNR 0540-69.3. The best-fit
power law model is presented as the histogram. The 
residuals (spectrum minus model) are shown in the lower panel.
\label{fig5}}
\end{figure}

\begin{figure} \caption{\asca SIS spectrum of N157B. The rest is the same
as Fig.~5.
\label{fig6}}
\end{figure}

\begin{figure}  \caption{Joint power law 
fit to the \asca and PSPC spectra of N157B. The rest is the same
as Fig.~5.
\label{fig7}}
\end{figure}

\begin{figure}  \caption{Joint 
fit to the \asca and PSPC spectra of N157B with a two-component model
consisting of a power law and a Gaussian line. The rest is the same
as Fig.~5.
\label{fig8}}
\end{figure}

\begin{figure} \caption{ \asca spectrum of N157B and the best joint-fit of
a power law to both the spectrum above 2~keV and the RHRI flux
of the comet-shaped X-ray feature (\S 3.3). The RHRI data point 
has its energy coverage between the 0.3-2.4~keV range.
The residuals of the fit are shown in the lower panel.
\label{fig9}}
\end{figure}

\begin{figure} \caption{Illustration of the current pulsar position and 
the flow of pulsar wind materials interacting with
the N157B remnant. The arrows represent the flow of the materials. The 
remnant's blast wave is represented by the outer shell. The
configuration of the components is somewhat similar to the fan-shaped radio 
feature connecting the pulsar PSR 1757-24 and 
the SNR G5.4-1.2 (Frail \& Kulkarni 1991).
\label{fig10}}
\end{figure}

\begin{references} 

\reference{} Angelini, L., et al. 1994, BAAS, 185, 6206
\reference{} Bessell, M. S. 1991, A\&A, 242, L17
\reference{} Burn, B. J. 1966, MNRAS, 133, 67
\reference{} Caplan, J., \& Deharveng, L. 1985, A\&AS, 62, 63
\reference{} Chu, Y.-H., R. C. Kennicutt, Jr., Schommer, R. A., Laff, J. 1992, 
AJ, 103, 1545 (CKSL92)
\reference{} Chu, Y-H. 1993, The Soft X-ray Cosmos, AIP Conf.
  Proc. 313, (AIP, New York), ed. E. M. Schlegel, \& R. Petre, p. 154
\reference{} Clark, D. H., et al. 1982, ApJ, 255, 440
\reference{} Cox, D. P. 1972, ApJ, 178, 159
\reference{} de Boer, K. S., Fitzpatrick, E. L., \& Savage, B. D. 1985, MNRAS, 217, 115
\reference{} Day, C. S. R., et al. 1995, ``The ABC Guide to \asca Data Reduction'', \asca Guest Observer Facility, NASA/GSFC
(http://heasarc.gsfc.nasa.gov/docs/software/ftools/ftools\_menu.html). 
\reference{} Dickel, J. R., et al. 1994, AJ, 107, 1067 (D94)
\reference{} Du Plessis, I. et al. 1995, ApJ, 453, 746
\reference{} Finley, J. P., \"Ogelman, H., Hasinger, G., \& Tr\"umper, J. 1993, ApJ, 410, 323
\reference{} Fitzpatrick, E. L. 1986, AJ, 92, 1068
\reference{} Frail, D. A., \& Kulkarni, S. R. 1991, Nature, 352, 785
\reference{} Gotthelf, E. V. 1993, CLEANSIS, available in the FTOOLS
software package (ftp legacy.gsfc.nasa.gov).
\reference{} Gotthelf, E. V., et al. 1997, in preparation
\reference{} Gould, A. 1995, ApJ, 452, 189
\reference{} Hughes, J. P., et al. 1995, ApJ, 444, L81
\reference{} Itoh, M., et al. 1994, New Horizon of X-ray Astronomy, ed. F. Makino, \& T. Ohashi, (Tokyo: Universal Academy Press), p507
\reference{} Jalota, L., Gotthelf, E. V., \& Zoonematkermani, S. 1993, SPIE, 4541, 453.
\reference{} Lyne, A. G., \& Graham-Smith, F. 1990, Pulsar Astronomy, Cambridge
University Press (Cambridge).
\reference{} Lyne, A. G., \& Lorimer, D. R. 1994, Nature, 369, 127
\reference{} Markwardt, C. B., \& \"Ogelman, H. 1995, Nature, 375, 40
\reference{} Mathewson, D. S., et al. 1983, ApJS, 51, 345
\reference{} Meyer, D. M., et al. 1994, ApJL, 437, 59
\reference{} Mills, B. Y., Turtle, A. J., \& Watkinson, A. 1978, MNRAS, 
185, 263
\reference{} Morrison, R., \& McCammon, D. 1983, \apj, 270, 119 
\reference{} Raymond J., \& Smith, B. W. 1994, XSPEC User's Guide for 
Version 8, NASA/GSFC.
\reference{} Russell, S. C., \& Dopita, M. A. 1992, ApJ, 384, 508
\reference{} Schild, H., \& Testor, G. 1992, A\&AS, 92, 729
\reference{} Seward, F. D., \& Harnden, Jr., F. R. 1994, ApJ, 421, 581
\reference{} Seward, F. D., \& Wang, Z.-R. 1988, ApJ, 332, 199
\reference{} Tanaka, Y., Inoue, H., \& Holt, S. S. 1994, PASJ, 46, L37
\reference{} Tr\"umper, J. 1983, Adv. Space Res., 2, 241
\reference{} Tucker, W. H. 1977, Radiation Processes in Astrophysics,
The MIT Press (Cambridge)
\reference{} Wang, Q. D. 1995, ApJ, 453, 783
\reference{} Wang, Q. D., \&  Helfand, D. J. 1991, ApJ, 370, 541
\reference{} Wang, Q. D., Li, Z.-Y., \& Begelman, M. C. 1993, Nature, 364, 127
\end{references}
\end{document}